\documentclass[conference]{IEEEtran}
\IEEEoverridecommandlockouts
\usepackage{cite}
\usepackage{amsmath,amssymb,amsfonts}
\usepackage{algorithmic}
\usepackage{graphicx}
\usepackage{textcomp}
\usepackage[table]{xcolor}
\def\BibTeX{{\rm B\kern-.05em{\sc i\kern-.025em b}\kern-.08em
    T\kern-.1667em\lower.7ex\hbox{E}\kern-.125emX}}
\usepackage{orcidlink}    
\usepackage{booktabs,multirow,bigstrut}
\usepackage[nameinlink,noabbrev,capitalise]{cleveref}
\usepackage{makecell}

\begin{document}

\title{QT-PUF: Quantum Tunneling Leakage Based PUF for Implantable IoMT Devices
\thanks{
\textsuperscript{*}Co-first authors. This work was supported by the National Research Foundation, Prime Minister’s Office, Singapore under its Campus for Research Excellence and Technological Enterprise (CREATE) - IN-CYPHER programme. \textsuperscript{\textdagger}Corresponding author email: v.mohan@imperial.ac.uk
}
}

\author{ 
Yueqi~Ma\textsuperscript{*1,2,3}~\orcidlink{0009-0002-3184-1403},
Vivek~Mohan\textsuperscript{*\textdagger1,2}~\orcidlink{0000-0002-0248-6417},
Chip Hong Chang\textsuperscript{2}~\orcidlink{0000-0001-6649-0573},
Emmanuel M. Drakakis\textsuperscript{1}~\orcidlink{0000-0001-6649-0573}

\\ \textsuperscript{1}\textit{Imperial College London, Imperial Global Singapore}~\\
\textsuperscript{2}\textit{Nanyang Technological University}~ \textsuperscript{3}\textit{Technical University of Munich}

}

\maketitle

\begin{abstract}
The Internet of Medical Things (IoMT) marks a shift toward decentralized healthcare, enabling continuous monitoring and personalized care through connected wearable and implantable devices. However, ensuring the trust and integrity of these devices themselves remains a major challenge, as physical compromise or counterfeiting can directly endanger patient safety, privacy, and data integrity. This work presents QT-PUF, a gate-tunneling-leakage-based physical unclonable function (PUF) that leverages quantum-mechanical gate leakage resulting from process-induced variations in standard CMOS devices. A differential readout circuit with a pseudo-resistor I-to-V frontend is proposed to convert the picoampere-level leakage variations into digital responses. Unlike existing PUFs such as those based on memory, ring oscillators, or arbiters, which are less suitable for ultralow-power IoMT devices (due to additional circuitry, power overhead, or poor stability), QT-PUF requires no external excitation or stabilization and operates under static bias. Simulation-based measurements for a $\mathbf{65}$~nm CMOS process demonstrate an entropy of $\mathbf{0.9999998}$, an FHD of $\mathbf{0.5001}$, and an average power (energy) consumption of $\mathbf{96.04}$~nW/bit ($\mathbf{19.21}$~fJ/bit, respectively) at $\mathbf{1.2\,V}$ and $\mathbf{35\,^{\circ}C}$ for the proposed PUF. It operates reliably across $\mathbf{0.9}\text{--}\mathbf{1.3}$~V and $\mathbf{0}\text{--}\mathbf{100\,^{\circ}C}$ with an average BER below $\mathbf{0.000163}$ across $\mathbf{1.0}\text{--}\mathbf{1.3}$~V and $\mathbf{10}\text{--}\mathbf{70\,^{\circ}C}$ within the operating conditions of typical implantable devices.
\end{abstract}

\vspace{-0.4cm}

\section{Introduction}
The Internet of Medical Things (IoMT) is reshaping modern healthcare by enabling continuous and decentralized monitoring of patients through wearable and implantable devices. These systems collect and transmit biomedical data in real time for remote diagnosis or on-device processing for therapeutic actions, reducing hospital loads and healthcare costs~\cite{Zhang2022LowCostAC, NCE2025, ISCAS2025_1, ISCAS2025_2, ISCAS2024, 10798270}. However, operating outside controlled clinical environments exposes sensitive physiological data to unauthorized access, potentially compromising patient safety and privacy~\cite{IoMT_Security_Advances, SEC-C-U, bracciale2023cybersecurity, Wu2021PrivacyPreservedEM}.

Conventional cryptographic mechanisms, relying on stored keys, are often vulnerable to invasive, non-invasive, and side-channel attacks~\cite{ZhengYue_Thesis}, and are unsuitable for ultra-low-power implantable electronics. Physical Unclonable Functions (PUFs) have emerged as lightweight alternatives, leveraging inherent manufacturing variations to generate unique, tamper-resistant responses without stored keys, enabling low-power, on-demand authentication and key generation~\cite{ZhengYue_Thesis}. Existing PUF designs, including arbiter, ring oscillator (RO), SRAM, optical, quantum, and hybrid approaches~\cite{GassendSiliconPR, Wang2019ASP, Babaei2019PhysicalUF, pappu2002physical, Kim2022RevisitingSA, Li2024QuantumPU, Moradi2017PhysicalUF}, face limitations that make them unsuitable for IoMT devices. For example, arbiter and RO PUFs exhibit high dynamic power consumption \cite{GassendSiliconPR,Wang2019ASP}; SRAM PUFs rely on reading the initial SRAM state at each power-on, but their reliability degrades after long dormancy or power loss and they remain susceptible to environmental fluctuations \cite{Babaei2019PhysicalUF}; while Optical and Quantum PUFs require bulky external components incompatible with compact implantable systems \cite{pappu2002physical, Kim2022RevisitingSA,Li2024QuantumPU}.

Leakage-based PUFs are promising for IoMT security due to their static operation, ultra-low power, and CMOS compatibility, with several designs demonstrating good stability\cite{lee2020354f,IOTJ_PUF,JSSC2019,Yuanfeng}. However, subthreshold-leakage-based PUFs \cite{lee2020354f,Yuanfeng} suffer from high temperature sensitivity, limited process randomness, and predictable leakage–parameter relationships that compromise uniqueness and modeling resistance, owing to thermally activated leakage that increases exponentially with temperature. In contrast, gate-tunneling-leakage PUFs offer high stability, entropy, and energy efficiency. The tunneling current, resulting from quantum-mechanical carrier transmission through the gate oxide, exhibits minimal temperature dependence~\cite{Lundgren1996TemperatureDC, Lenzlinger1969FowlerNordheimTI, Yassine1997TemperatureDO}, while its strong nonlinearity with oxide thickness, doping, and interface traps enhances uniqueness and modeling resistance. Operating at picoampere-to-nanoampere static bias, these PUFs enable ultra-low-power, stable operation ideal for implantable IoMT devices.  

\begin{figure*}[t!]
\centering
\includegraphics[width=0.9\linewidth]{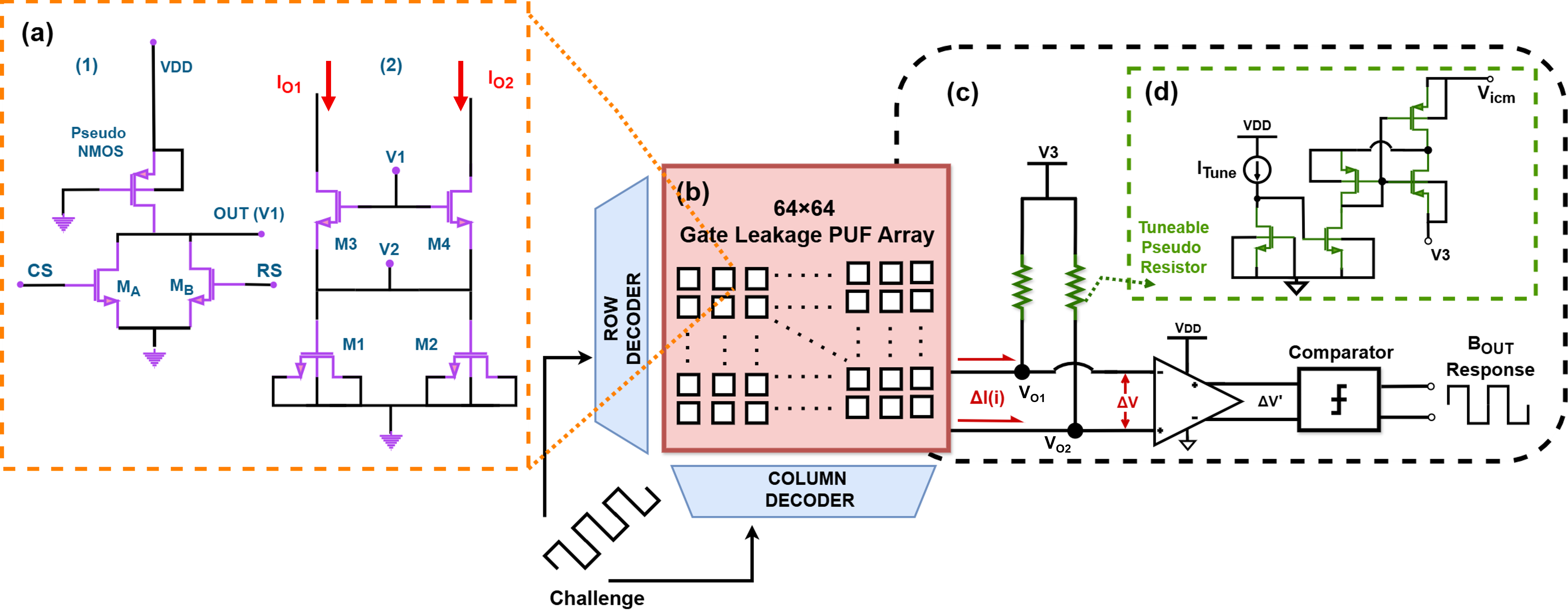}
\caption{Schematic of QT-PUF system architecture composed of: (a-1) cell selection logic (CS and RS are the column and row select signals) and (a-2) QT-PUF Cell that are arranged into a (b) 4096-bit QT-PUF Array, and (c) readout circuit featuring (d) a pair of pseudo-resistor for I to V conversion, followed by amplifier and comparator stages to convert the intrinsic quantum gate leakage variations into digital response ($\mathrm{B_{OUT}}$).
\vspace{-0.5cm}
}
\label{fig:PUFsys}
\end{figure*}

In this work we introduce a PUF implemented in standard CMOS, paired with a differential readout circuit suitable for IoMT applications, and make the following contributions: 
\begin{itemize}
    \item We propose QT-PUF, a gate-tunneling-leakage array PUF composed of four-transistor cells, where two larger transistors generate measurable quantum gate leakage currents, providing ultra-low-power ($\mathrm{96.04~nW/bit}$), static, and stable operation for implantable IoMT devices.   
    \item We present a pseudoresistor-based I-V converter, and differential amplifier readout circuit that reliably converts gate-leakage variations into digital responses via a comparator circuit with high entropy ($0.9999998$) and low BER ($0.000163$).
    \item We validate the overall design through simulations in $65$ nm CMOS process, demonstrating strong uniqueness and robustness across voltage and temperature variations, as reflected in the inter-array FHD of $0.5001$.  
\end{itemize}

\section{Methodology}
This section outlines the principles of quantum gate-tunneling leakage as an entropy source, the PUF cell and array design, readout circuitry, and challenge–response (CRP) generation. QT-PUF (Fig.~\ref{fig:PUFsys}) integrates five functional modules that collectively convert physical randomness into a digital response, implemented in standard $\mathrm{65~nm}$ CMOS and optimized for ultra-low-power, stable operation in implantable IoMT devices.
\vspace{-0.2cm}

\subsection{Gate Leakage}
\begin{figure}[t]
\centering
\includegraphics[width=0.9\linewidth]{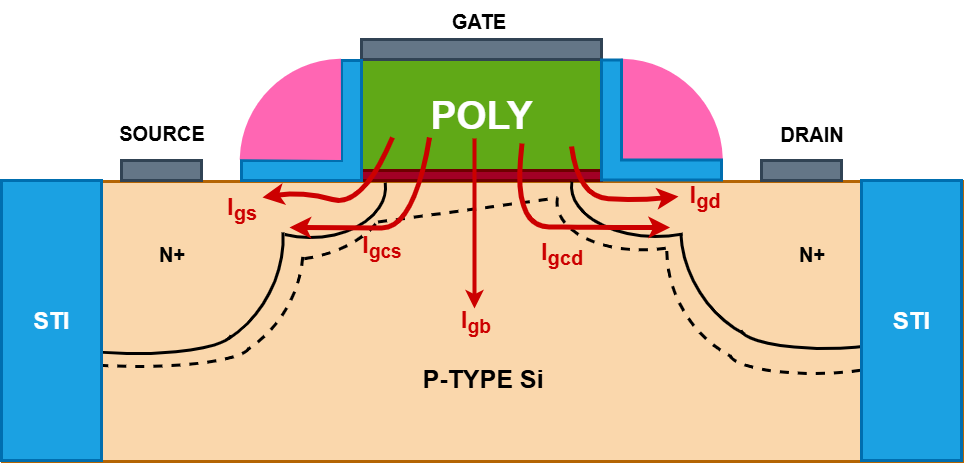}
\caption{Schematic illustrating the five gate leakage components in NMOS devices with ultrathin gate oxide layers: $\mathrm{I_{gs},~I_{gd},~I_{gcs},~I_{gcd}}$, and $\mathrm{I_{gb}}$. STI: Shallow Trench Isolation.
\vspace{-0.5cm}
}
\label{fig:DeviceLeakage}
\end{figure}

%\vspace{-0.1cm}
The primary gate-leakage mechanisms in MOS devices with ultrathin silicon dioxide are Fowler–Nordheim tunneling (FNT), quantum mechanical direct tunneling (QMDT), and trap-assisted tunneling (TAT)~\cite{Lo1997QuantummechanicalMO,Shih1998ModelingGL,Chaudhry2013FundamentalsON,Cao2000BSIM4GL}. Net gate leakage is the superposition of these mechanisms: FNT dominates at high gate voltages, QMDT in thin oxides under normal operation, and TAT under low-bias or high-defect conditions, as shown in Fig.~\ref{fig:DeviceLeakage}. 
Ultra-thin gate oxides give rise to five main leakage components: gate-source ($\mathrm{I_{gs}}$), gate-drain ($\mathrm{I_{gd}}$), gate-channel-source ($\mathrm{I_{gcs}}$), gate-channel-drain ($\mathrm{I_{gcd}}$), and gate-substrate ($\mathrm{I_{gb}}$) currents~\cite{ranuarez2006review,Shih1998ModelingGL,Nagano1994MechanismOL}.

\subsection{QT-PUF Cell Design}
In this work, QMDT is exploited as the primary entropy source and therefore, in our proposed PUF cell design, a voltage $V_2\approx$ $\mathrm{1.2}$~V is applied to the gate terminal of the tunneling device, lower than typical FNT ($\mathrm{>3~V}$) and higher than biasing conditions for TAT.
The fundamental building block of the proposed PUF system is the PUF cell designed as shown Fig.~\ref{fig:PUFsys}(a-2). It is composed of two NMOS transistors (M1, M2) whose drain, source, and substrate are grounded, thereby acting as a gate leakage sources, and control transistors (M3 and M4) in each branch which allow the leakage current to flow to outputs $\mathrm{I_{O1}}$ and $\mathrm{I_{O2}}$. 
When the gate bias of M1 and M2 exceeds approximately $0.9$ V, an inversion layer is established near the Si–SiO$\mathrm{2}$ interface, leading to quantum confinement of carriers that travel laterally along the interface with minimal substrate injection~\cite{Roy2003LeakageCM, Shih1998ModelingGL}. Due to limited lateral diffusion overlap, $\mathrm{I_{gs}}$ and $\mathrm{I_{gd}}$ remain small, whereas $\mathrm{I_{gcs}}$ and $\mathrm{I_{gcd}}$ dominate the total gate tunneling current. These leakage variations provide a stable, unique and unclonable entropy source for each cell.

\subsection{QT-PUF Array Architecture}
As shown in Fig.~\ref{fig:PUFsys}(b), the system core comprises a $\mathrm{64\times64}$ PUF cell array ($\mathrm{4,096}$ cells in total), each exhibiting a unique leakage pattern from process-induced variations such as oxide thickness and doping differences. An applied challenge is decoded into row and column signals via row and column address decoders, respectively, to select target cells (via V1) while suppressing interference from unselected cells. The picoampere level leakages of the selected cells ($\mathrm{I_{O1}}$ and $\mathrm{I_{O2}}$) are routed to a readout circuit that converts them into stable digital response. Each cell’s distinct leakage ensures high uniqueness and entropy across the array.

\begin{figure}[t!]
\centering
\includegraphics[width=1.0\linewidth]{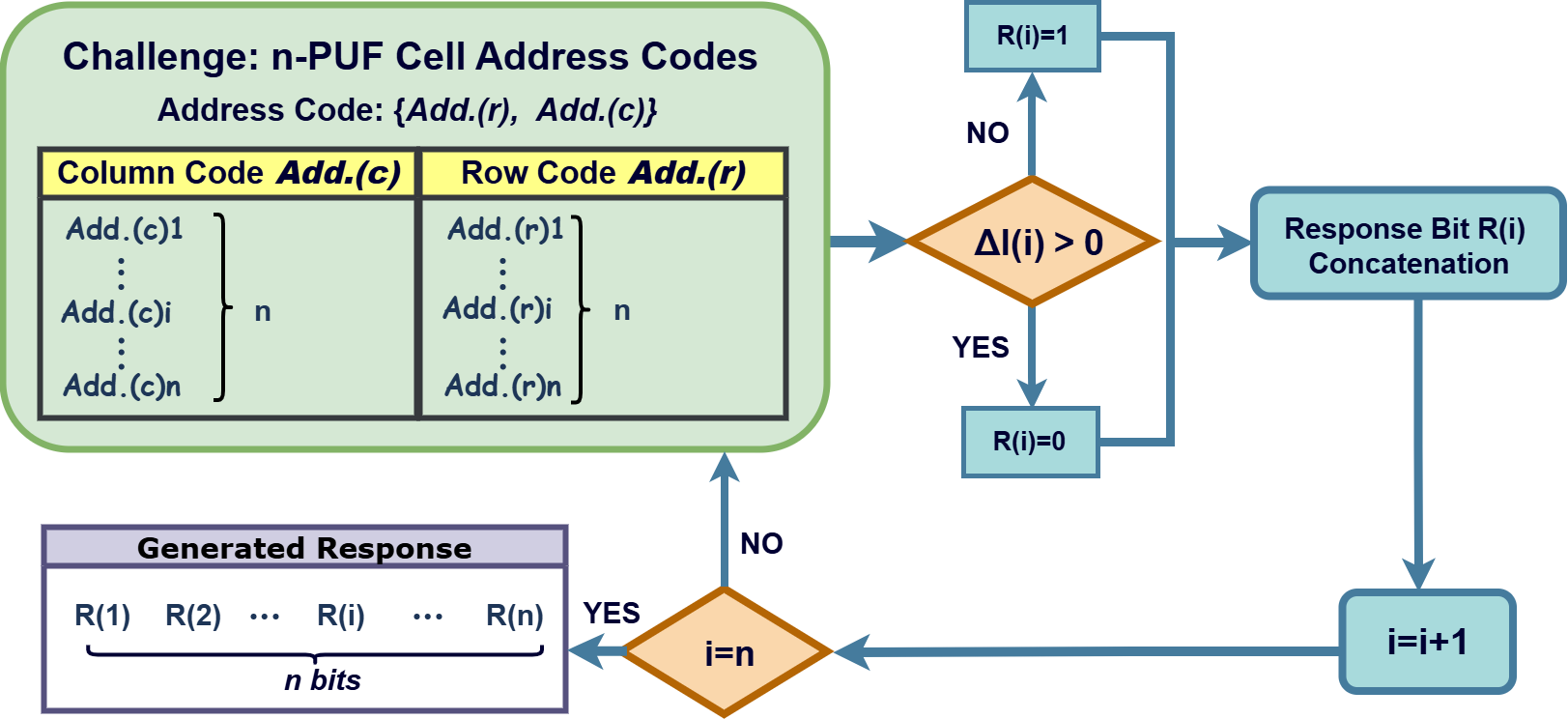}
\caption{CRP generation workflow: a challenge with address and response length $n$ triggers current comparisons in selected cells to generate $n$ concatenated response bits.
\vspace{-0.6cm}}
\label{fig:CRP}
\end{figure}
\subsection{Address Decoding and Cell Selection}
Row and column decoders convert input address codes of a challenge into row (RS) and column (CS) select signals. The selection logic, shown in Fig.~\ref{fig:PUFsys}(a-1), ensures only the targeted cell contributes to the readout, preventing crosstalk. Pseudo-NMOS devices in the selection circuit control the differential input pair $\mathrm{M_A}$ and $\mathrm{M_B}$ to enable or disable current paths for each cell.

\subsection{Readout Circuitry}
The readout circuit converts the picoampere-level leakage current from the selected cell into a stable digital bit. First, a pair of tunable pseudo-resistors (shown in Fig. \ref{fig:PUFsys}(d)) transform the leakage current ($\mathrm{I_{O1}}$ and $\mathrm{I_{O2}}$) into a voltage according to the relation: $\mathrm{V_{O1(O2)} = V_{3} - I_{O1(O2)} \times R_{pseudo}}$, where $\mathrm{V_{3}}$ is the biasing voltage, and $\mathrm{R_{pseudo}}$ is the effective resistance (in the order of G$\Omega$). This converted differential voltage ($\mathrm{\Delta V = V_{O1}-V_{O2}}$) is then amplified by a fully differential operational amplifier with a gain of $30$~dB to suppress noise and ensure signal integrity (resulting in $\mathrm{\Delta V_{}'}$), before being digitized by a comparator to produce a stable and unique output bit ($\mathrm{B_{out}}$) as shown in Fig. \ref{fig:PUFsys}(c).

\subsection{Challenge–Response Pair (CRP) Generation}
A challenge \emph{C} consists of $n$ address codes corresponding to the desired response length `$n$'. For each address code (split into row and column addresses) of a challenge \emph{C}, the readout circuit compares the voltage-converted leakage currents corresponding to the two branches of the selected cell. The comparison result is latched as a 1-bit response (0 or 1), where the dominant branch determines the output state. Concatenating $n$ such response bits for $n$ address codes of a challenge results in an $n$-bit PUF response \emph{R} as illustrated in Fig.~\ref{fig:CRP}, which is repeatable for the same challenge and PUF unit, unique and unclonable for each PUF unit.

\section{Results}
The proposed gate-leakage PUF is prototyped in a $\mathrm{65}$~nm CMOS process. We first verified and evaluated a small-scale PUF array before scaling to a $\mathrm{4096}$-cell PUF. A comparison with state-of-the-art designs is summarized in Table \ref{tab1}. 

\subsection{16-cell PUF Array Test}
\begin{figure}
    \centering
    \includegraphics[width=0.99\linewidth]{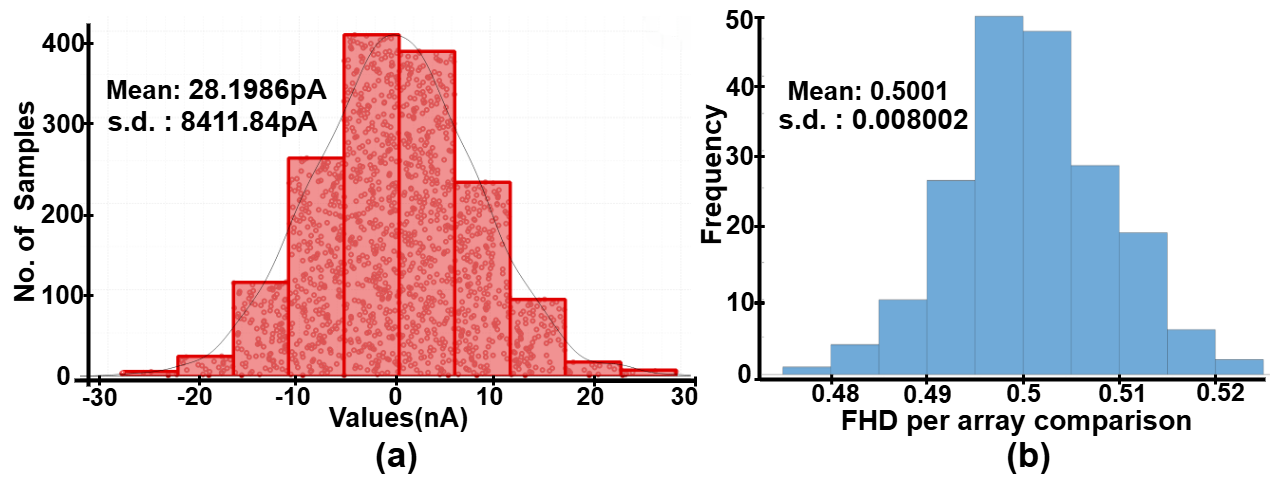}
    \vspace{-0.6cm}
    \caption{(a) Distribution of differential leakage currents: the mean value is $28.1986~$pA, and the standard deviation is $8411.84~$pA; (b) Uniqueness is evaluated over 190 inter-array comparisons generated by 20 arrays $(19+18+...+2+1)$ using 4096 bits of data each, yielding an average FHD of 0.5001 and a standard deviation of 0.008002.}
    \vspace{-0.4cm}
    \label{fig:FHD}
\end{figure}

Monte Carlo simulations were first performed on a single cell to evaluate process-induced variations. As shown in Fig.~\ref{fig:FHD}(a), the distribution of differential leakage currents exhibits a mean of $28.1986$~pA and a standard deviation of $8.41184$~nA, validating the expected quantum tunneling based leakage current variability.

\begin{figure}
    \centering
    \includegraphics[width=0.95\linewidth]{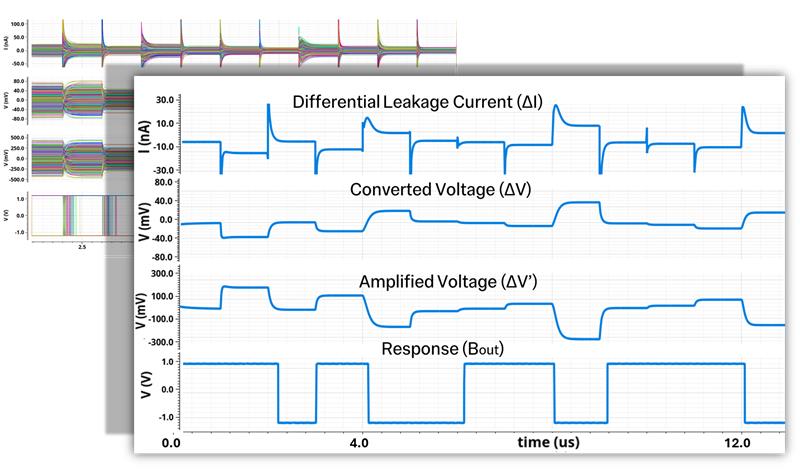}
    \caption{PUF waveforms of 1500 samples obtained from simulation with one highlighted as an example. Due to process-related randomness, the 1500 differential leakage current waveforms are all different. The input challenges are applied at the rate of $1$~MHz.}
    \vspace{-0.3cm}
    \label{fig:16cells_MC_waves}
\end{figure}

For array-level measurement, 16 PUF cells were sampled over 1500 cycles each. Representative waveforms are illustrated in Fig.~\ref{fig:16cells_MC_waves}. Due to random oxide thickness and channel variations, the differential leakage current (first row) may appear as positive or negative across cells. This signal is processed through a current comparator to produce a binary output bit (last row). The intermediate voltage waveform after current-to-voltage conversion and filtering (second row) effectively suppresses noise and transient glitches before amplification (third row), ensuring bit stability across samples.

%\vspace{-0.4cm}

\begin{figure}[t!]
    \centering
    \includegraphics[width=0.99\linewidth]{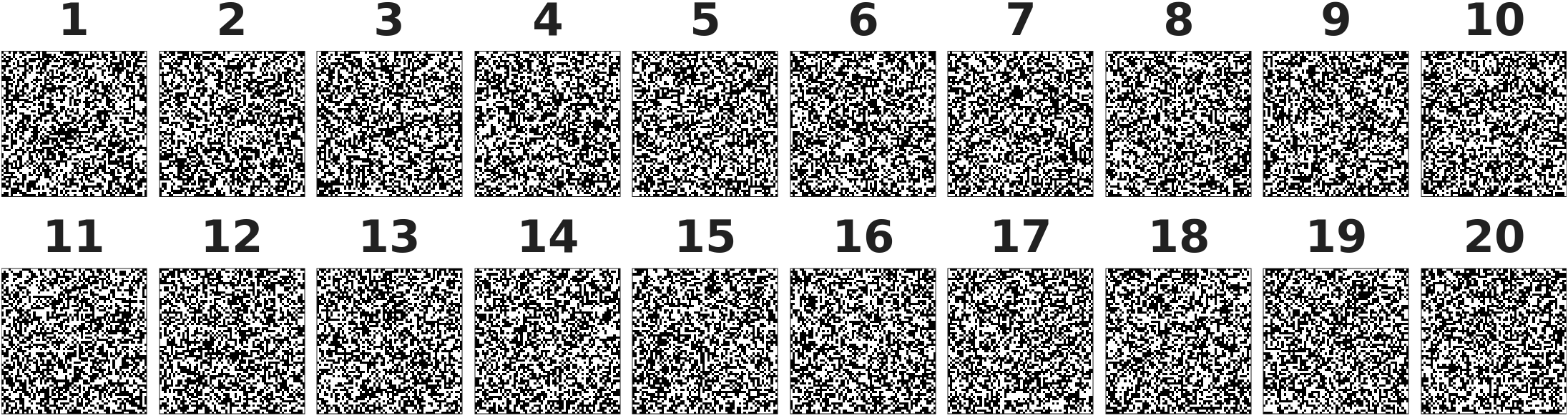}
    \caption{Logical speckle figure of measured PUF output from 20 arrays: Black[1]-50.02\%, White[0]-49.98\%
    \vspace{-0.3cm}}
    \label{fig:speckle}
\end{figure}

\begin{table}[t!]
%\vspace{-0.45cm}
\caption{Results of NIST Randomness Tests on 700 Arrays}
\begin{tabular}{|l|c|c|}
\hline \textbf{Test} & \textbf{Average p-value} & \textbf{Pass Rate (\%)} \\
\hline Monobit Test & 0.7524 & 100\\
\hline Frequency in a Block Test & 0.5356 & 99.43\\
\hline Runs Test &  0.4886 & 98.57\\
\hline Longest Run of 1s in a Block Test & 0.5579 & 99.43\\
\hline DFT Test & 0.4956 & 98.43\\
\hline Serial Test &  0.9996 & 100\\
\hline NIST Approximate Entropy Test & 0.5886 & 99.43\\
\hline Cumulative Sums Test  & 0.9899 & 100\\
\hline
\end{tabular}
\label{tab:NIST}
\vspace{-0.4cm}
\end{table}

\begin{table*}[tp!]
\centering
\caption{Comparison of the proposed QT-PUF with the state of the art}
\resizebox{\linewidth}{!}{
\begin{tabular}{|c|c|c|c|c|c|c|c|c|}
\hline
\textbf{Reference} & 
\textbf{Process (nm)} & 
\textbf{Topology} & 
\textbf{Entropy} & 
\makecell{\textbf{Area per}\\\textbf{bit ($\mathbf{F^2}$)}} & 
\makecell{\textbf{Stabilization}\\\textbf{Required}} & 
\makecell{\textbf{Inter-}\\\textbf{Array FHD}} & 
\textbf{BER (\%)} & 
\makecell{\textbf{Power/Energy}\\\textbf{per bit}} \\
\hline
JSSC 2019 \cite{JSSC2019} & 40 & Soft Oxide Breakdown & 0.9998 & 1875 & NO & 0.496 & 0.02 & 51.2 fJ \\
\hline
ISSCC 2020 \cite{Choi2020274PU} & 28 (FDSOI) & NAND & 0.9999999 & 3698.97 & YES & 0.4978 & 0.089 / 1.41e-64$^{h}$ & 0.38 mW \\
\hline
JSSC 2020 \cite{Liang2020AWV} & 28 & Ring Oscillator & 0.99986 & 33163 & YES & 0.4994 & 0.156 & 2150 fJ \\
\hline
JSSC 2021 \cite{lee2020354f} & 180 & Sub-threshold Leakage & N/G$^a$ & 354 & YES & 0.4946 & 0.0073 & 465000 fJ \\
\hline
TCAS 2022 \cite{9737288} & 65 & SRAM & 0.9999965 & 307.69 & NO & 0.4947 & 3 & 0.324 mW / 81 fJ \\
\hline
TCAS 2023 \cite{9942277} & 28 & Arbiter & 0.9999927 & 5143 & YES & 0.502 & 0.0016 & 1.486 mW \\
\hline
IOTJ 2024 \cite{IOTJ_PUF} & 55 & Gate-oxide Leakage & N/G$^a$ & 3239.67 & NO & 0.500 & 3 & N/G$^a$ \\
\hline
Nature 2025 \cite{Liu2025ChipscaleRC} & N/A$^{\mathrm{b}}$ & Carbon Nanotubes & 0.7–0.95 & N/G$^a$ & YES & 0.48 & 0.55–0.98 & N/G$^a$ \\
\hline
TCAS 2025 \cite{10836795} & 45 & SRAM & N/G$^a$ & 46419.75$^d$ & YES & 0.4937–0.4989 & 0 & $\propto N^{e}$ \\
\hline
ISSCC 2025 \cite{10904785} & 28 & Eye-Opening Arbiter & 0.99906 & 44438.78 & YES & 0.4999 & 3.49 / 2e-6$^{h}$ & 259 fJ \\
\hline
\textbf{QT-PUF (This work)} & 65 & Gate Tunneling Leakage & 0.9999998 & 9151.95 & NO & 0.5001 & 0.0163 & 96.04 nW / 19.21 fJ \\
\hline
\multicolumn{9}{l}{
\begin{minipage}[t]{0.98\textwidth}
\footnotesize
N/G$^a$: Information not given in the paper;\,
N/A$^b$: Not Applicable;\,
$^{d}$ 4 CRPs;\,
$^{e}$ Increasing \#challenges ($N$) raises overall consumption;\,
$^{h}$ After stabilization.
% \vspace{-0.4cm}
\end{minipage}
}
\end{tabular}}
\label{tab1}
\end{table*}

\subsection{64×64 PUF Array Evaluation}
A $64\times64$ PUF array was further tested to assess large-scale randomness and uniqueness. The logical speckle pattern in Fig.~\ref{fig:speckle} demonstrates a near-ideal balance between ‘0’ and ‘1’ responses, with probabilities of 0.4998 and 0.5002, respectively. The corresponding Shannon entropy, calculated using Eq.~\ref{shannon}, is 0.9999998, approaching the theoretical maximum for random binary sequences.
\vspace{-0.1cm}
\begin{equation}
E=-\left[p{\log }_{2}p+\left(1-p\right){\log }_{2}\left(1-p\right)\right]
\label{shannon}
\end{equation}

The uniqueness of twenty arrays is evaluated using inter-array fractional Hamming distance (FHD), calculated across $190$ PUF array/unit pairs (each with $\mathrm{4096}$ bits response). The histogram in Fig.~\ref{fig:FHD}(b) shows a mean FHD of $0.5001$ and a standard deviation of $0.008002$, indicating high uniqueness and independence among arrays. Randomness was further verified using the NIST SP 800-22 suite across $700$ arrays as summarized in Table~\ref{tab:NIST}, along with array pass percentage and corresponding mean p-values for each sub-test. Statistical analysis predicts that at a significance level of $0.001$, at least $99.41\%$ of arrays will satisfy all sub-test criteria.

\begin{figure}
    \centering
    \includegraphics[width=0.98\linewidth]{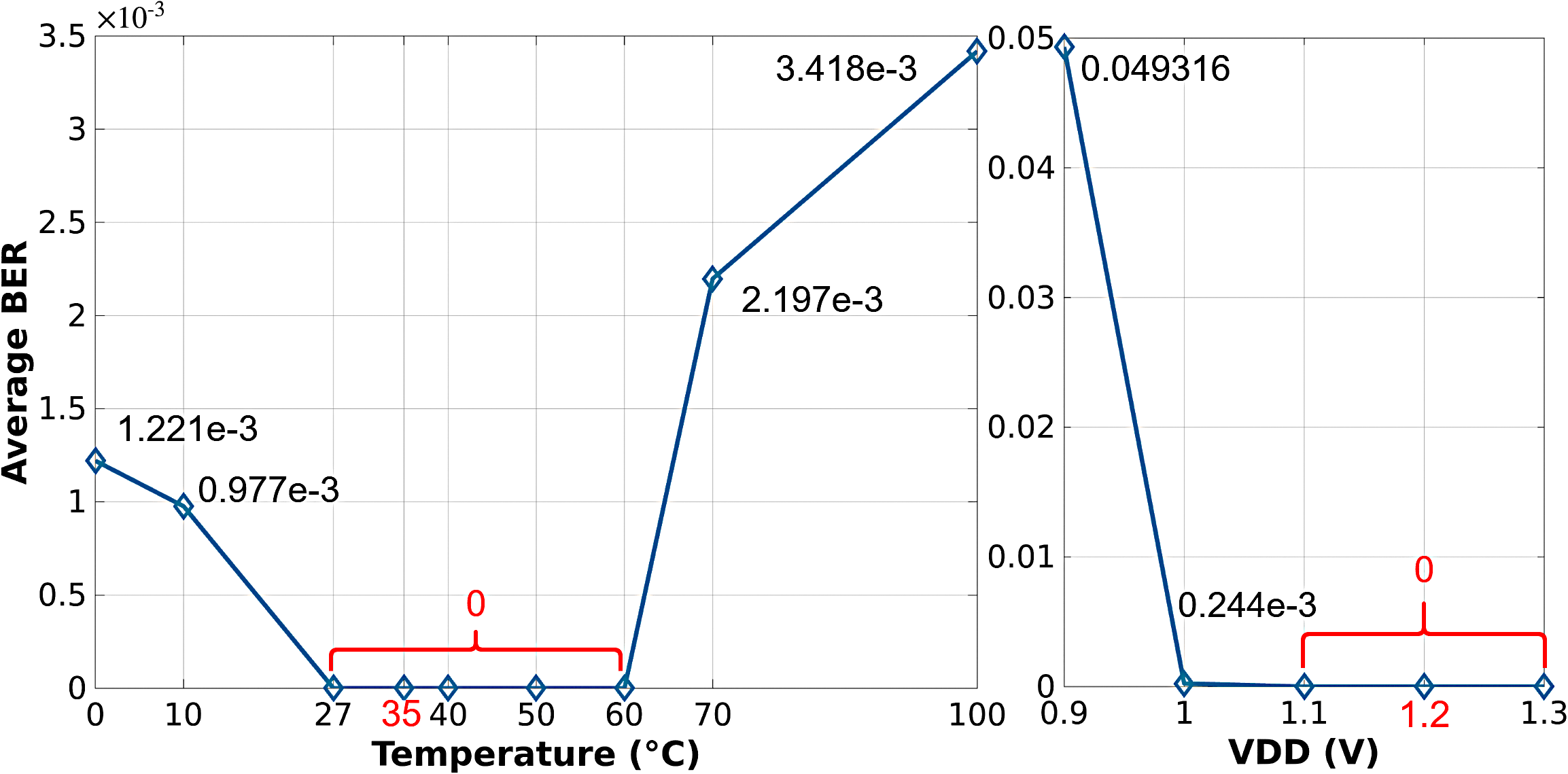}
    \caption{Measured BER of the proposed PUF under (a) temperature variation and (b) supply voltage (VDD) variation. The standard operating condition is 1.2 V and 35$^{\circ}\mathrm{C}$.
    \vspace{-0.6cm}}
    \label{fig:BER}
\end{figure}

Reliability under voltage and temperature variations is shown in Fig.~\ref{fig:BER}. 
The worst-case BER occurs at $100^{\circ}\mathrm{C}$ and $1.2\,\mathrm{V}$ ($0.003418$), 
and at $0.9\,\mathrm{V}$ and $35^{\circ}\mathrm{C}$ ($0.049316$). 
%%%%%%%%%%%%%%%%%%%%%%%%%%%
 Averaged over the range of $10$--$70^{\circ}\mathrm{C}$ and $1.0$--$1.3\,\mathrm{V}$, which sufficiently covers typical implantable IoMT operating conditions, the BER remains below $0.000163$.
 %%%%%%%%%%%%%%%%%%%%%%%%%%%%%%
The average power consumption per bit is $96.038\,\mathrm{nW}$ at $1.2\,\mathrm{V}$ and $35^{\circ}\mathrm{C}$, 
making the proposed design suitable for ultra-low-power embedded and implantable operation.

% \vspace{-0.4cm}
\subsection{Comparison and Discussion}
Table \ref{tab1} summarizes a comparison between the proposed gate-tunneling-leakage PUF and prior state-of-the-art designs. The proposed PUF achieves a near-ideal entropy of $0.9999998$ and an inter-array FHD of $0.5001$, demonstrating excellent randomness and uniqueness. In contrast to many stabilization-based implementations \cite{Choi2020274PU, Liang2020AWV, lee2020354f, 9942277, Liu2025ChipscaleRC, 10836795, 10904785}, the proposed design operates reliably without requiring additional calibration or error correction. Although its area per cell is slightly larger than that of deep-submicron SRAM or NAND PUFs, the circuit features a simple structure and inherently stable operation. Moreover, it attains a low BER of \textcolor{black}{$\mathrm{0.000163}$} without any post-processing and a power (energy) consumption of $\mathrm{96.038}$~nW/bit ($\mathrm{19.207}$~fJ/bit, respectively), surpassing most recent CMOS-based PUF implementations. Overall, the proposed PUF achieves a favorable balance among reliability, uniqueness, and energy efficiency, making it well-suited for ultra-low-power implantable/wearable IoMT applications.
\vspace{-0.2cm}

\section{Conclusion}
This work presents a quantum tunneling leakage-based PUF that leverages intrinsic gate leakage in standard CMOS for hardware security in implantable IoMT devices. By exploiting the nonlinear and temperature-insensitive characteristics of quantum mechanical tunneling, the proposed design achieves high entropy ($0.9999998$), excellent uniqueness (FHD = $0.5001$), and low BER ($0.000163$) under operating conditions of typical implantable devices without additional stabilization or error correction. Using $65\,\mathrm{nm}$ CMOS, it operates reliably across voltage and temperature ranges ($0.9$--$1.3\,\mathrm{V}$ and $0$--$100^{\circ}\mathrm{C}$) with an ultra-low power cost of \textcolor{black}{$96.038\,\mathrm{nW/bit}$}, outperforming many existing CMOS PUFs.
These results demonstrate that gate-tunneling leakage can serve as a robust, low-power, and scalable entropy source, offering a promising pathway towards secure and energy-efficient hardware primitives for next-generation implantable IoMT systems. Future work will focus on optimizing the QT-PUF cell, performing hardware measurements, and exploring security applications such as mutual authentication protocols utilizing QT-PUF integrated on IoMT devices.

%\clearpage % Forces all floats to be processed and placed

\bibliographystyle{IEEEtran}
\bibliography{references}
\end{document}